\documentclass[]{spie}  
\usepackage[]{graphicx}

\include{00README.XXX}

\title{Medical Application Studies at ELI-NP} 

\author{D. Habs\supit{1}, P.G. Thirolf\supit{1}, C. Lang\supit{1}, 
 M. Jentschel\supit{2}, U. K\"oster\supit{2}, F. Negoita\supit{3} 
and V. Zamfir\supit{3}
\skiplinehalf
\supit{1}  Fakult\"at f\"ur Physik, 
       Ludwig-Maximilians Universit\"at M\"unchen, D-85748 Garching, Germany\\
\supit{2} Institut Laue Langevin, 6 rue Jules Horowitz, 
           F-38042 Grenoble Cedex 9, France\\
 \supit{3} National Institute of Physics and Nuclear Engineering(IFIN-HH), 
          Bucharest-Magurele 077125, Romania   
}

\authorinfo{Further author information: (Send correspondence to D.~Habs)\\
D.~Habs: E-mail: dietrich.habs@lmu.de, Telephone: 0049 89 2891 4077}

\begin{document} 
  \maketitle 

\begin{abstract}
 
We study the production of radioisotopes for nuclear medicine 
in ($\gamma,\gamma'$) photoexcitation reactions or
$(\gamma,x{\rm n}+y{\rm p})$ photonuclear reactions  
for the examples of $^{195m}$Pt,$^{117m}$Sn and $^{44}$Ti 
with high flux [($10^{13}-10^{15}$)$\gamma$/s], small beam diameter 
and small energy band width ($\Delta E/E \approx 10^{-3}-10^{-4}$) $\gamma$ beams. 
In order to realize an optimum $\gamma$-focal spot, a refractive 
$\gamma$-lens consisting of
a stack of many concave micro-lenses will be used. It allows for the 
production of a high specific activity and the use of enriched isotopes.
For photonuclear reactions with a narrow $\gamma$ beam, the energy 
deposition in the target can be reduced by using a stack of thin target 
wires, hence avoiding direct stopping of the Compton electrons and 
$e^+e^-$ pairs. The well-defined 
initial excitation energy of the compound nucleus leads to a small 
number of reaction channels and enables new combinations of target
isotope and final radioisotope. The narrow-bandwidth $\gamma$ 
excitation may make use of collective resonances in $\gamma$-width,
leading to increased cross sections. $(\gamma,\gamma')$ 
isomer production via specially selected $\gamma$ cascades allows to produce 
high specific activity in multiple excitations, where no back-pumping of 
the isomer to the ground state occurs. 
The produced isotopes will open the way for completely 
new clinical applications of radioisotopes. 
For example $^{195m}$Pt 
could be used to verify the patient's response to chemotherapy 
with platinum compounds before a complete treatment is performed.
In targeted radionuclide therapy the short-range Auger and conversion
electrons of $^{195m}$Pt and $^{117m}$Sn enable a very local treatment.
The generator $^{44}$Ti allows for a PET with an additional $\gamma$-quantum
($\gamma$-PET), resulting in a reduced dose or better spatial resolution.
\end{abstract}


\keywords{ Gamma-ray sources, Photonuclear reactions, Radioactive sources 
for therapy, Nuclear medicine imaging, Radiopharmaceuticals, Gamma-ray lens}

\section{INTRODUCTION}
\label{sec:intro}
This publication is intended to focus in more detail on three 
medical radioisotpes $^{195m}$Pt, $^{117m}$Sn and $^{44}$Ti, while 
the more general concepts of medical radioisotope production with 
brilliant $\gamma$-beams have been introduced in Ref.~\cite{habskoester10}. 
In nuclear medicine radioisotopes are used for diagnostic and 
therapeutic purposes \cite{schiepers06,cook06}. Many diagnostics 
applications are based on molecular imaging methods, i.e. either 
on positron emitters for 3D imaging with PET (positron emission 
tomography) or gamma ray emitters for 2D imaging with planar 
gamma cameras or 3D imaging with SPECT (single-photon emission 
computer tomography)\footnote{Today the nuclear medicine imaging 
techniques PET and SPECT (using radionuclides injected into 
the patient's body) are frequently combined in the same apparatus 
with the radiology technique CT (computer tomography) to PET/CT 
or SPECT/CT, respectively}. 
The main advantage of nuclear medicine methods is the
high sensitivity of the detection systems that allows using tracers at
extremely low concentrations. This extremely low amount of radiotracers assures
that they do not show any (bio-)chemical effect on the organism. Thus, the
diagnostic procedure does not interfere with the normal body functions and
provides direct information on them, not perturbed 
by the detection method. To maintain these 
intrinsic advantages of nuclear medicine diagnostics,
one has to assure that radiotracers of {\it relatively high 
specific activity} are used, i.e. that the injected radiotracer 
is not accompanied by too many stable isotopes of the same 
(or a chemically similar) element.

Radioisotopes are also used for therapeutic applications, in 
particular for endo-radiotherapy. Targeted systemic therapies 
allow fighting diseases that are non-localized, e.g. leukemia and 
other cancer types in an advanced state, when already multiple 
metastases have been created. Usually a bioconjugate 
\cite{schiepers06,Bos07} is used that shows a 
high affinity and selectivity to bind to peptide receptors 
or antigens that are overexpressed 
on certain cancer cells with respect to normal cells. 
Combining such a bioconjugate with a suitable radioisotope, such as a
low-energy electron emitter, allows irradiating and destroying
selectively the cancer cells. Depending on the nature of the bioconjugate,
these therapies are called Peptide Receptor Radio Therapy (PRRT) 
\cite{cook06,Reu06} when peptides are used as bioconjugates or 
radioimmunotherapy (RIT) \cite{cook06,jac10}, when 
antibodies are used as bioconjugates. 
The radioisotopes for labeling of the bioconjugates should 
have a high specific activity to minimize injection of bioconjugates 
labeled with stable isotopes that do not show radiotherapeutic efficiency. 
Thus again {\it high specific activities} are required for radioisotopes used 
in such therapies. In the cases of $^{195m}$Pt and $^{117m}$Sn for 
radiotherapeutic applications a low-energy $\gamma$ transition can be used 
for SPEC imaging.

The long-lived $^{44}$Ti isotope ( T$_{1/2}\approx$ 60a) will be used as a 
generator to produce the shorter-lived $^{44}$Sc (T$_{1/2}$= 4 h), which has
the unique property to populate predominantly the first excited 1157 keV 
state in $^{44}$Ca. Thus the $e^+$ annihilation into two counterpropagating 
511 keV  $\gamma$-quanta is accompanied by the third prompt $\gamma$ transition 
of 1157 keV, allowing a unique localization of each individual decay
\cite{Grignon07}. Previously already one $^{44}$Ti generator has been produced 
via the $^{44}$Sc(p,2n) reaction in an irradiation for several months.
All the development of the $^{44}$Ti/$^{44}$Sc generator
and the medical application of $^{44}$Sc to patients have been established
\cite{roescha10,roeschb10}.

The radioisotopes for diagnostic or therapeutic nuclear medicine applications
are usually produced by nuclear reactions. The required projectiles are
typically either neutrons from dedicated reactors or charged
particles from cyclotrons or other accelerators. In our publication 
\cite{habskoester10} we have compared these methods in detail
to a production scheme using a $\gamma$-beam and discussed a large variety of
new possible radioisotopes produced by $\gamma$-beams. In section 2 
we describe the most recent developments on brilliant $\gamma$-beams. 
In section 3 we describe a $\gamma$ lens, with which the specific activity 
can be increased by optimum focusing at the production target. While the 
$\gamma$ facilities allow to produce many radioisotopes 
in new photonuclear reactions with significantly higher specific activity, 
we here focus on the three radioisotopes $^{195m}$Pt, $^{117m}$Sn and $^{44}$Ti.
In section 4 we discuss these special cases.


\section{Highly brilliant $\gamma$ sources}

The new $\gamma$ beams are produced via Compton back-scattering of
laser photons from a relativistic electron beam.
The presently world-leading facility for photonuclear physics is the 
High-Intensity $\gamma$-ray Source (HI$\gamma$S) at Duke University (USA).
It uses the Compton back-scattering of photons, provided by a high-intensity 
Free-Electron Laser (FEL), in order to produce a brilliant $\gamma$ beam.
The $\gamma$ intensity in the energy range between 1\,MeV and 160\,MeV amounts 
to $10^{8}s^{-1}$ with a band width of about 5$\%$~\cite{weller09}. The 
brilliant Mono-Energetic Gamma-ray (MEGa-ray) facility at Lawrence Livermore 
National Laboratory (USA) is based on a normal-conducting electron linac and 
will yield already in 2012 for $\gamma$ energies between 0.5 and 2.5 MeV a 
$\gamma$ intensity of $10^{13}s^{-1}$ with an energy band width of down to 
$10^{-3}$ and a brilliance of 
$10^{22}/[mm^2mrad^2s 0.1\%BW]$~\cite{Bar10megaray}. Using the same
accelerator technology, at the upcoming Extreme Light Infrastructure -
Nuclear Physics (ELI-NP) facility in Bucharest, until 2015 a $\gamma$ beam will 
become available, providing about the same $\gamma$ intensity 
and band width in the 
energy range of 1-19\,MeV~\cite{ELINP10}. \\
At present, great efforts are also invested all over the world 
to realize highly brilliant $\gamma$ beams based on the Energy 
Recovery Linac (ERL) technology. The Energy Recovery Linac (ERL) 
requires a new type of superconducting electron accelerator that provides 
a highly brilliant, high-intensity electron beam. 
The main components of an ERL are an electron injector, a superconducting 
linac, and an energy recovery loop. 
After injection from a highly brilliant electron source, the electrons 
are accelerated by the time-varying radio-frequency field of the 
superconducting linac. The electron bunches are transported once through 
a recirculation loop and are re-injected
into the linac during the decelerating RF phase of the superconducting cavities.
So the beam dump has to take the electron bunches only with low energy, while
the main part of the electron energy is recycled. At ERL's, high-energy, 
highly brilliant $\gamma$ beams can be created by Compton 
back-scattering of photons with high energy (0.1-100\,MeV), again 
recirculating the photons in a very high finesse cavity with MW power
to overcome the small Compton cross section.
ERL technology is pioneered at Cornell University (together with 
Thomas Jefferson National Laboratory)~\cite{Bil10a,Bil10b,liepe10},
where an ERL is presently constructed for a 5\,GeV, 100 mA electron beam.
At the KEK accelerator facility in Japan, an ERL project is presently pursued
aiming at a $\gamma$ beam with an intensity of 
$10^{13}s^{-1}$~\cite{Haj08,Hay10}. In Germany, a high-current and 
low-emittance demonstrator ERL facility (BERLinPro) 
is developed at the Helmholtz Zentrum Berlin~\cite{Jan10}.
There are mainly three different operation modes of an ERL: high-flux mode, 
high-brilliance mode, and a short-pulse 
mode~\cite{Bil10a,Bil10b,Hay10}. For our purpose of medical isotope 
production the high brilliance mode is of particular interest. 
The ultimately envisaged photon intensity is $>10^{15}s^{-1}$ in an 
energy range of 0.5 - 25\,MeV. Such a facility would provide a brilliant 
pulsed (ps pulse length) $\gamma$ beam with a narrow band width of 
about $< 10^{-3}$, and a low emittance of $10^{-4}mm^2mrad^2$.  

\section{$\gamma$ lens with monochromatisation}

\begin{figure}
   \begin{center}
   \begin{tabular}{c}
   \includegraphics[height=4cm]{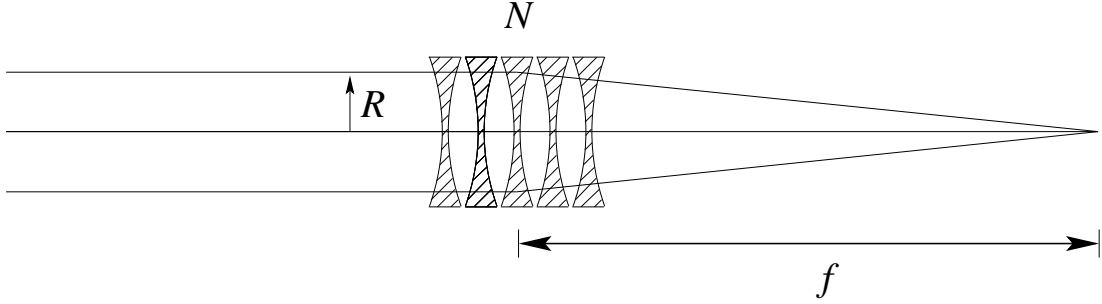}
   \end{tabular}
   \end{center}
   \caption[lens] 
   { \label{fig:lens} $\gamma$-lens consisting of N micro-lenses, irradiated 
 with a $\gamma$-beam of radius $R$. The $\gamma$ lens has a focal length $f$.}
   \end{figure}

Lenses for hard X-rays are well established 
\cite{yang93,snigirev98,schroer05,schroer10} for energies up to
200 keV. At synchrotrons the brilliance decreases more than exponential
above 100 keV energies. Here we want to extend the range of
these lenses to MeV-$\gamma$-beams, making use of their many orders
of magnitude higher brilliance. Since the real part $\delta$ of the index of 
refraction $n = 1-\delta + i\beta$ decreases with $1/E_{\gamma}^2$, the 
realization of sufficent focusing strength becomes more difficult. 
On the other hand, the imaginary part $\beta$ of the index of refraction 
decreases even faster with $1/E_{\gamma}^{3.5}$. At higher energies new processes
like Compton scattering and pair creation contribute to $\beta$. Only with the 
new highly brilliant $\gamma$-beams the smaller beam radius $R$ allows to reach 
reasonable focal lengths $f$. A further trick to reduce the focal length is 
to stack a larger number $N$ of micro-lenses behind each other.
The focal length $f$ is given by
\begin{equation}
  \label{eq:lens1}
        f=\frac{R}{2\delta N}
\end{equation}

For the real part $\delta$ of the index of refraction we have:

\begin{equation}
  \label{eq:lens2}
     \delta=\frac{r_e \lambda_{\gamma}^2}{2\pi} \rho \frac{Z}{A}
       \propto \frac{1}{E_{\gamma}^2}
\end{equation}

Here $r_e$ is the classical electron radius, 
$\lambda_{\gamma}=\frac{\hbar c}{E_{\gamma}}$ is the $\gamma$ wavelength, 
$\rho$ is the lens density and Z and
A are the charge and mass number of the lens material, respectively. 
Since the real part $\delta$ of the index of refraction is negative 
for $\gamma$-energies, we show concave lenses in Fig. \ref{fig:lens}.

We now have to consider the special case that we will use in 
combination with the $\gamma$-lens always $\gamma$-beams produced
by Compton back-scattering from relativistic electrons.
Here the $\gamma$-energy $E_{\gamma}$ is given by:

\begin{equation}
  \label{eq:compton}
     E_{\gamma}=\frac{2\gamma_e^2 \cdot (1+\cos{\phi}) \cdot E_L}{1+(\gamma_e\Theta)^2+\frac{4\gamma_e E_L}{m_ec^2}}
\end{equation}

$E_L$ is energy of the primary laser photons, $\gamma_e=\frac{E_e}{m_ec^2}$
is the $\gamma$-factor of the electron beam with energy $E_e$ and the
electron rest mass $m_e$ and we consider that the laser beam enters under
an angle $\phi$ with respect to the electron direction and the outcoming
$\gamma$ photon has an angle $\Theta$ with respect to the electron beam 
direction. Thus for $\Theta=1/\gamma_e$ the emitted $\gamma$ photon has
about a factor of 2 smaller $\gamma$-energy that the $\gamma$-photon in
electron beam direction. For angles 
$\Theta_{BW}\le \frac{\sqrt{BW}}{\gamma_e}$,
the central $\gamma$ beam opening cone, it is no longer the $\Theta$
dependance of the Compton scattering, but other factors like the band width
of the electron beam or the band width of the laser beam, which determine
the band width BW of the $\gamma$-beam. We will now limit the acceptance
angle of the $\gamma$ lens or the corresponding $\gamma$ beam radius
$R$ at the entrace to the lens to this much smaller angle $\Theta_{BW}$,
where furthermore the main intensity of the $\gamma$ beam is concentrated.
Thus we make use of the very special $\gamma$ energy angle correlation
of Compton backscattered $\gamma$ rays to work  with a very small
radius $R$ and to obtain a natural monochromatisation. The outer angles 
$\Theta$ are no longer focused by the lens. We can experimentally monitor
the energy range focused by the lens by using a special nuclear resonance
$\gamma$ energy and measure the nuclear resonance fluorescence (NRF)
for a thin wire target, probing the $\gamma$ beam focus behind the 
$\gamma$ lens. The better the band width of the central $\gamma$ beam 
the smaller the radius $R$ and the focal length $f$ will be. For higher
$\gamma$ beam energies and correspondingly larger $\gamma_e$ 
the radius $R$ will be smaller, partially compensating the strong
$\gamma$ energy dependance of the focal length $f$ with          
$\propto \frac{1}{E_{\gamma}^2}\propto \frac{1}{\gamma_e^4}$.
By increasing the number $N$ of micro-lenses
and using materials of higher density $\rho$, we can keep the focal
length sufficiently small for $\gamma$ beams with sufficiently good
primary band width BW. In the very central region the very thin 
micro-lenses will lead to small absortion of the $\gamma$ beam.
We have to align the $\gamma$ lens very accurately with respect to 
the electron beam direction, but again this can be monitored
by NRF measurements. 

When reducing the $\gamma$ beam radius along the
stack of lenses, the shape of the lenses can be adjusted adiabatically 
and very small focal spots can be reached \cite{schroer05}. 
For our production of medical radioisotopes we want to reach a very small 
spot size on the target to obtain higher specific activity and to allow 
for the use of isotopically enriched targets. The small diameter wire-targets 
allow for an easy escape of Compton electrons and pairs, resulting in 
little heating.

\section{Special Radioisotopes}

\subsection{The nuclei $^{195}$Pt and $^{117}$Sn} 

$^{195}$Pt and $^{117,119}$Sn have many common properties.
$^{195}$Pt has above its $1/2^-$[530] Nilsson ground state a 
$13/2^+$[606] isomer at 259 keV with a halflife of 4~d. The isomer
originates from the $1i_{13/2}$ intruder state from the next higher shell. 
For the semi-magic $^{117,119}$Sn with Z=50, the situation is very similar
to $^{195}$Pt, except that all angular momenta and the main shell quantum
number are reduced by one. Thus the intruder $1i_{13/2}$ state of $^{195}$Pt
is replaced by the intruder $1h_{11/2}$ state for $^{117}$Sn. 
The ground state 
$1/2^+$[404] of $^{117,119}$Sn originates from a 2d$_{3/2}$ state. 
The $11/2^-$ isomer at 315 keV of $^{117}$Sn originates from the 
1h$_{11/2}$ neutron state and has a half-life of $ T_{1/2}$=~13.6 d.
The corresponding $11/2^-$ isomer of $^{119}$Sn at 89.5 keV has
a half-life of $T_{1/2}$=293 d, which is too long for medical applications.

But also the neighbouring even-even nuclei have many similarities.
The Pt nuclei have close-lying shell 
model states with $\Delta N=1,\Delta l=\Delta j =3$ \cite{butler93,zamfir89}, 
here the $i_{13/2}$ and $f_{7/2}$ states, which lead to octupole deformation
and low-lying collective $3^-$ states. These $3^-$ octupole states are 
coupled to $2^+$ quadrupole states, resulting in collective $1^-$ states.
One could also couple a collective $5^-$ and $4^+$ to a collective $1^-$ 
state. These $1^-$ states have rather large dipole moments of about 0.1 e fm.
The $5^-$ state may be due to shell model states with an angular 
momentum difference $\Delta l=\Delta j=5$ of the $i_{13/2}$ and $p_{3/2}$ states.
In this way one could explain the strong E1 core excitation and its 
strong collective core deexcitation by a $1^-\rightarrow 3^-\rightarrow 5^-$
cascade with collective E2 transitions. The $5^-$ core excitation then will
decay predominantly to the i$_{13/2}^+$ state. 
The f and p shell-model states lying close to the i shell-model state
explain the collective $3^-,5^-$ core excitations in the neighbouring 
$^{194,196}$Pt isotopes. Also on top of the $13/2^-$ isomer low-lying collective
quadrupole excitations exist, which will mix with the high-spin states
of the core $3^-$ and $5^-$ excitations and the ${1/2}^+$ ground state, leading
to the strong cascade coupling the gateway state with the  $1^-$  
core excitation to the high spin isomer.

Also for the Sn nuclei we have close-lying shell-model states with
$\Delta N=1,\Delta l=\Delta J =3$ to explain the strong octupole 
state and by coupling to the $2^+$ quadrupole state the dipole E1 
transition \cite{butler93}.

We can extrapolate all these collective states of neighbouring even-even 
nuclei to the odd-neutron $^{195}$Pt and $^{117}$Sn. We verify a 
weak coupling of the core excitations for known collective states in 
these odd nuclei. Thus we predict a $1/2^+$ and $3/2^+$ E1 excitation from 
the $1/2^-$ ground state
to the $1^-$ core excitation at about 1.8 MeV for $^{195}$Pt and a
$1/2^-$ and $3/2^-$ E1 core excitation from the ground state of $^{117}$Sn
at about 3.8 MeV.  

\subsubsection{Production of the Isomers $^{195m}$Pt and 
$^{117m}$Sn}

\begin{figure}
   \begin{center}
   \begin{tabular}{c}
   \includegraphics[height=8cm]{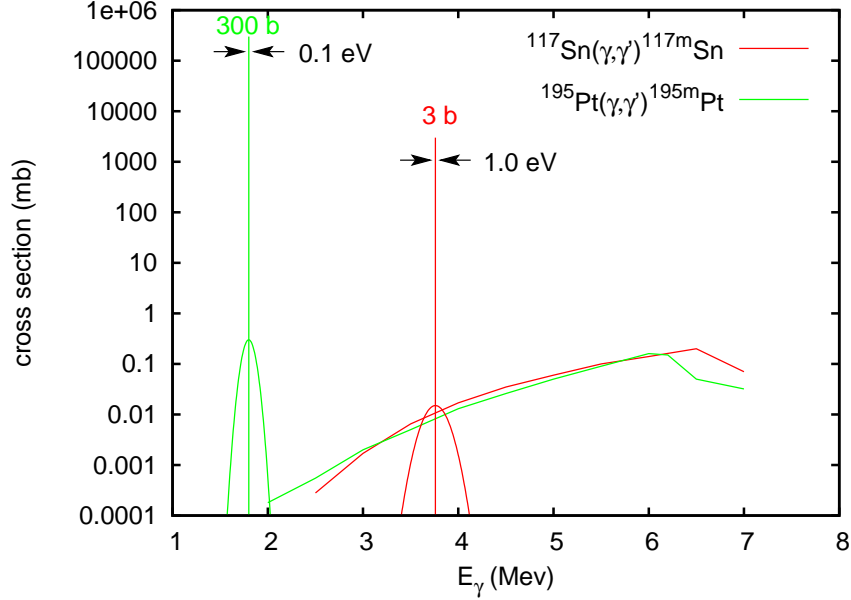}
   \end{tabular}
   \end{center}
   \caption[Pt+Sn] 
   { \label{fig:Pt+Sn} Statistical model calculation for the cross section
    with the TALYS code, giving the broad cross-section dependence and 
   predicted additional collective resonances deduced from neighbouring 
   even-even nuclei.}
   \end{figure}

In Fig. \ref{fig:Pt+Sn} we show the predicted photonuclear cross-sections as 
a function of $\gamma$-energy. Besides the statistical calculations with the 
TALYS code and standard parameters, we show the new expected collective 
gateway states. Their relative yield has to be adjusted to reproduce
the integrated yields.  
At present for $^{195}$Pt and $^{117}$Sn only one survey experiment 
including 19 nuclides exists, where the integrated cross section
$\sigma\cdot\Gamma$ was deduced for one bremsstrahlung spectrum ranging
from 0.5 MeV to 4 MeV and another one ranging from 1.0 MeV to 6 MeV. 
A {\it ``three to four orders of magnitude larger integrated cross section
than usually''} was observed including $^{195m}$Pt and
$^{117m}$Sn~\cite{car91}. In the analysis one 
single gateway state at 2.125 MeV was artificially introduced
for all 19 nuclides, when deducing the effective $\sigma\cdot \Gamma$. 
We now want to perform at HI$\gamma$S a search for these individual resonant
gateway states for $^{195}$Pt and $^{117}$Sn. Then we could localize
them more accurately with the better band width of the MEGa-ray or ELI-NP
facilities. These resonances would facilitate the
photonuclear production of these isomers significantly.

\subsubsection{Medical Applications of $^{195m}$Pt and $^{117m}$Sn} 

\paragraph{$^{195m}$Pt: Determining the efficiency of chemotherapy 
for tumors and the optimum dose by nuclear imaging}
In chemotherapy of tumors most often platinum cytotoxic compounds like 
cisplatin or carbonplatin are used. We want to label these compounds with 
$^{195m}$Pt for pharmacokinetic studies like tumor uptake and want to exclude
``nonresponding'' patients from unnecessary chemotherapy and optimizing 
the dose of all chemotherapy. For such a diagnostics a large-scale market
can be foreseen, but it would also save many people from painful treatments.
We estimated in Ref.~\cite{habskoester10} that several hundred patient-specific 
uptake samples per day could be produced with a $\gamma$ beam facility
if optimum gateway states 
are identified by scanning the isomer production with high $\gamma$ beam 
resolution. $^{195m}$Pt has a high 13/2$^+$ spin isomer at 259 keV, halflife
(T$_{1/2}$= 4 d) with SPECT transitions of 130 keV and 99 keV to the 1/2$^-$  
ground state.

\paragraph{$^{195m}$Pt: Cancer Therapy with Short-Range Electrons}

\begin{figure}[t!]
   \begin{center}
   \begin{tabular}{c}
   \includegraphics[height=3cm]{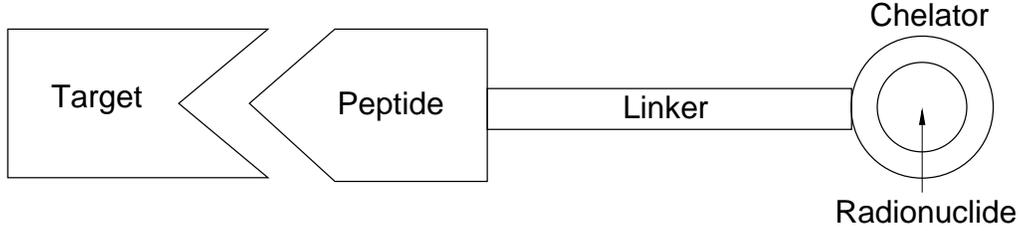}
   \end{tabular}
   \end{center}
   \caption[PRRT] 
   { \label{fig:PRRT} 
Schematic picture explaining how a specific radionuclide
            can be transported to a specific petide receptor.}
   \end{figure}

On the other hand, $^{195m}$Pt may be used in cancer therapy by transporting
it to the tumor by specific bioconjugates \cite{Dow00}. 
Special antibodies may be used 
in radio-imunotherapy (RIT) or special peptides in Peptide Receptor 
Radionuclide therapy (PRRT) to guide the radioactive isotope to the tumor. 
In Fig. \ref{fig:PRRT} we show how a radionuclide like $^{195m}$Pt is 
transported with a chelator to the specific peptide receptor.

 Here $^{195m}$Pt is of high interest,
because the emitted conversion electrons and Auger electrons have a 
rather small range of 1.5-200 $\mu$m and low electron energies of 
7 keV, 17 keV, 21 keV and 116 keV, respectively. They allow to 
kill locally small tumor 
clusters, while the dose given to normal tissue is rather low. Several new
therapeutic radioisotopes with this small range and high LET can be produced 
by $\gamma$ beams.

\paragraph{$^{117m}$Sn An emitter of low energy Auger electrons 
for tumor therapy}

Auger electron therapy requires targeting into individual tumor cells, 
even into the nucleus or to DNA, due to the short range below 1$\mu$m
of the Auger electrons; but there it is of high REB due to the shower 
of the abundantly produced 5-30 Auger electrons~\cite{Buc06}. 
On the other hand, Auger radiation is of low toxicity, while being 
transported through the body. Thus Auger electron therapy \cite{Bis00}
needs special tumor-specific transport molecules like antibodies or peptides.
Many of the low-lying high spin isomers produced in $(\gamma,\gamma')$ 
reactions have strongly converted transitions, which trigger these
large showers of Auger cascades.

Once the optimum production of the isomers $^{195m}$Pt and $^{117m}$Sn
has been established, the scientific focus will shift to produce the
best bioconjugates to transport these radioisotopes to specific cancer cells.
\begin{figure}[b]
   \begin{center}
   \begin{tabular}{c}
   \includegraphics[height=3.5cm]{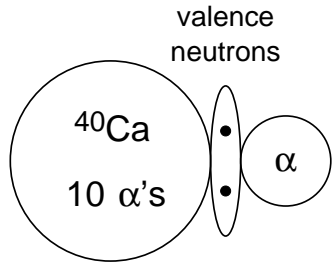}
   \end{tabular}
   \end{center}
   \caption[44Ti0] 
   { \label{fig:44Ti0} $\alpha$-cluster model of $^{46}$Ti.}
   \end{figure}

\subsection{The Generator $^{44}$Ti}

\subsubsection{Photoproduction of $^{44}$Ti}

In the $\alpha$-cluster model, $^{46}$Ti and $^{44}$Ti have a simple structure. 
Both nuclei have a strongly bound doubly magic $^{40}$Ca core, which in the
$\alpha$-cluster model consists of 10 $\alpha$ particles forming the spherical
core. For $^{44}$Ti we add an additional $\alpha$ particle to the doubly
magic $^{40}$Ca core, obtaining a deformed nucleus. $^{46}$Ti can be obtained
by adding two valence neutrons to the $^{44}$Ti. This structure is shown in
Fig. \ref{fig:44Ti0}.  

\begin{figure}[t!]
   \begin{center}
   \begin{tabular}{c}
   \includegraphics[height=7cm]{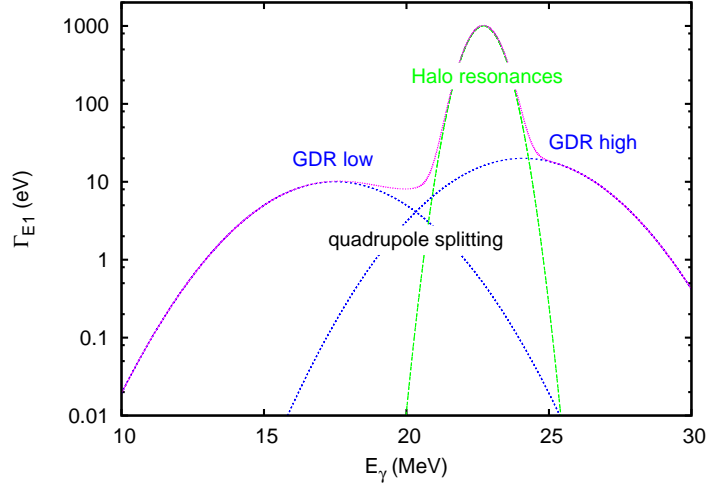}
   \end{tabular}
   \end{center}
   \caption[44Ti1] 
   { \label{fig:44Ti1} Schematic E1 strength of $^{46}$Ti. The $\gamma$ width
 of about 1.3 eV at 13.2 MeV is confirmed in Ref.~\cite{guttormsen11}.}
   \end{figure} 
The very strongly bound $\alpha$ particles make it probable that we obtain 
two loosely bound halo neutrons when we photo-excite $^{46}$Ti, such that it
 can emit just two low-energy neutrons. Since these halo neutrons result in 
large dipole monents and strong E1 excitations, we expect strong collective
resonances close to the reaction threshold. 
We expect a deformation splitting with a deformation of $^{46}$Ti 
of about $\beta=0.3$, resulting in a high-energy component
$E_{GDR \perp}=E_{GDR}\cdot (1+\beta/3)$, corresponding to about 24.1 MeV, 
and a low-energy component $E_{GDR z}=   E_{GDR}\cdot (1-\beta\cdot (2/3))$, 
corresponding to about 17.5 MeV. This GDR is shown in Fig. \ref{fig:44Ti1} 
with its splitting due to the quadrupole 
deformation into a high-energy component $E_{GDR \perp}$ with an 
oscillation normal to the z axis, and a component half as strong 
at $E_{GDR z}$ along the z axis.
Microscopically, in the E1 excitation we have to lift up two 1f7/2 neutrons 
to form $1^-$ states. Thus we will have $3s_{1/2}$ neutron levels populated 
close to zero binding energy, but also some odd angular 
momentum states. These weakly bound states will decay sequencially into 
two neutrons above the 2n-threshold, but due to the s-neutron components 
in the wavefunctions they will have extended wave functions
and an enhanced E1 strength.

\begin{figure}[h!]
   \begin{center}
   \begin{tabular}{c}
   \includegraphics[height=7cm]{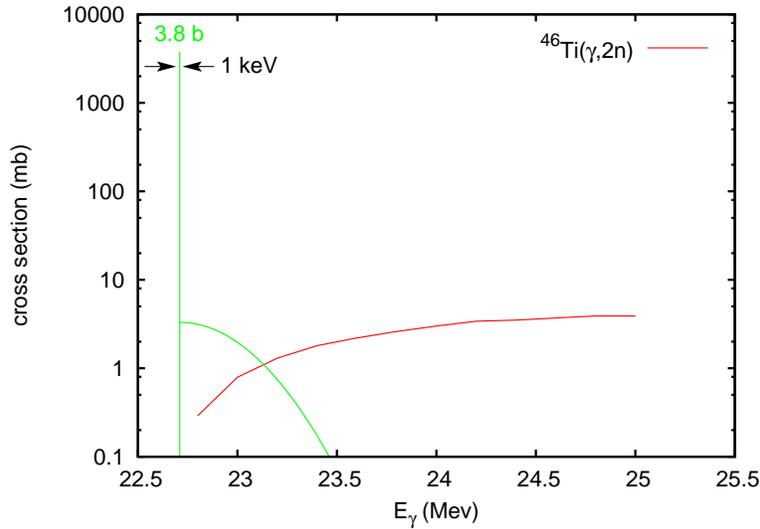}
   \end{tabular}
   \end{center}
   \caption[44Ti2] 
   { \label{fig:44Ti2} Cross section for the $^{46}$Ti($\gamma,2n$)$^{44}$Ti
 reaction calculated with the TALYS code. Also shown is a halo resonance
at the reaction threshold.}
   \end{figure} 

In Fig. \ref{fig:44Ti2} we show in red the usual energy dependence of 
the $^{46}$Ti($\gamma$,2n)$^{44}$Ti
cross section from statistical model simulations with the code TALYS. In
addition we hope to find this predicted strong new resonance. 
      
\begin{figure}[h]
   \begin{center}
   \begin{tabular}{c}
  \includegraphics[height=6.5cm]{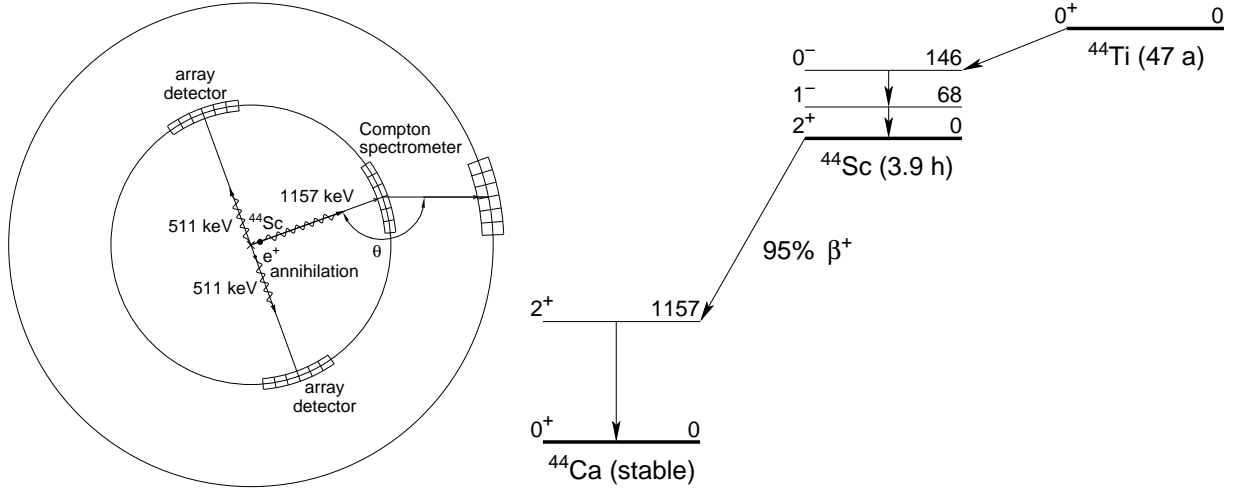}
   \end{tabular}
   \end{center}
   \caption[44Ti3] 
   { \label{fig:44Ti3} Schematic picture for the combined $\gamma$-PET and
                         the $^{44}$Ti/$^{44}$Sc generator.}
   \end{figure} 

\subsubsection{A $^{44}Ti \rightarrow ^{44}$Sc generator for the 
               $\gamma$-PET isotope}

In recent years PET (Positron Emission Tomography) with a typical spatial
resolution of 3-10 mm was supplemented by multi-slice X-ray CT
(Computer Tomography) of much better resolution, leading to the novel
technology of PET/CT. The co-registration and the reference frame of 
CT are very helpful for the interpretation of PET images. CT and PET require a
comparable dose. Looking at such images it is very apparent, that a significant 
improvement in the resolution of PET is highly desirable. $^{44}$Sc is the 
best candidate to supply the two 511 keV annihilation quanta together 
with a strongly populated 1.157 MeV transition as shown in 
Fig. \ref{fig:44Ti3}. By measuring the position and direction of this 
$\gamma$-ray accurately with a Compton spectrometer together with the two 
511 keV quanta, the location of the emitting nucleus can be located in 3 
dimensions. In conventional PET the two collinear 511 keV quanta only 
allow to determine a two-dimensional localization on a line of response. 
Thus a much better
spatial resolution can be achieved for the same dose with $\gamma$-PET compared
to PET. With $^{44}$Ti (half-life $T_{1/2}$= 59 a), the production 
with a very promising generator for $^{44}$Sc ($T_{1/2}$= 3.9 h) becomes 
available with $\gamma$ beams. Again a much stronger population via the
fine structure of the Giant Dipole Resonance (GDR) 
in the $^{46}$Ti($\gamma$,2n)$^{44}$Ti reaction is expected for 
the $^{44}$Ti core consisting of the doubly magic $^{40}$Ca
and an $\alpha$ particle. The long halflife of $^{44}$Ti requires a large
transmutation with an intense $\gamma$ beam, but on the other hand leads to
a very valuable, long-lived generator. The production of $^{44}$Ti in the
$(\gamma$,2n) reaction requires rather large $\gamma$ energies of 21-24 MeV,
and would require an increase of the maximum presently planned $\gamma$ 
energy at ELI-NP of 19 MeV. This seems possible, since an increase of
the electron beam energy by 10\% should be achievable after some 
experience with the accelerator cavities. 

\subsubsection{$\gamma$-PET}

The Medium Energy Gamma-ray Astronomy library (MEGAlib)\cite{megalib06}
contains all software tools to perform 
simulations for Compton cameras \cite{kanbach05,frandes10}. 
Similar to measurements in PET detectors, we are including here the 
detector positions of the coincident 511 keV $\gamma$-rays emitted 
back-to-back in order to determine their common direction. 
Then we combine this direction of the two 511 keV quanta  
with the direction of the Compton cone of the 1157 keV quantum.
Here are to possiblities of a Compton camera. In the first method
we only measure the energy of the Compton scattered electron in the
detector together with the direction and energy of the scattered $\gamma$ ray
and obtain by energy considerations the Compton cone as possible directions
of the primary $\gamma$ ray. In a more refind measurement we can
measure the momentum of the Compton electron, we measure also its direction.
Then we can limit the direction of the primary $\gamma$ photon to a small
section of the Compton cone. We reach accuracies for the primary $\gamma$ ray
of about $1^0$. 
In this way we can filter out on an event-by-event basis the good events, 
where the reconstruction of the 3D decay location is restricted very much.
For the lower energy 511 keV quanta we probably only can reconstruct
their Compton cones to determine some Compton scattering in the patient,
while for the 1257 keV $\gamma$ photons we can measure the Compton electron
momentum. This reconstruction of the source is very different from the usual
PET reconstruction, where the lines of different annihilation decays are 
combined to reconstruct the source position. If one or more of the
three $\gamma$ quanta (511 keV, 511 keV, 1157 keV) undergo a Compton scattering
in the patient, they are strongly suppressed by the analysis including
the algorithmus of the Compton camera, reducing 
the blurring of the usual PET-analysis by this Compton scattering within
the patient strongly. While the diffusion of the positron 
with a maximum energy of 2 MeV results in a deviation 
in the annihilation position from the original $^{44}$Sc decay, the prompt
1157 keV photon originates from the $^{44}$Sc decay. In detailed simulations
we want to determine in how far the $\gamma$-PET technique results in 
a strong dose reduction and improved spatial resolution compared 
to the classical PET analysis.   

This $\gamma$-PET imaging techniques can be tested with a $^{22}$Na source, 
where the $\beta^+$ decay again leads predominantly (in about 90\%) to the 
$2^+$-level of $^{22}$Ne at 1274.6 keV. However, the long lifetime of 
$^{22}$Na of 2.6 a allows no use in medicine.

\section{Conclusion}

Many further new interesting medical radioisotpes can be produced
(see Ref.\cite{habskoester10}): new 'matched pairs' of isotopes 
of the same element become available, one for diagnostics, the other 
for therapy, allowing to contol and optimize the transport of the
isotope by a bioconjugate to the tumor. Also new therapy isotopes 
become available like $^{225}$Ac, where four consecutive $\alpha$ decays 
can cause much more DNA double-strand breaking. 
Developing these techniques and applications is a promising 
task of ELI-NP with a strong societal component.


{\bf Acknowlegdement}
We thank A.~Zoglauer, G.~Kanbach and R.~Diehl to get started with 
the MEGAlib code package for simulating Compton cameras. We thank A.~Tonchev  
for helping us with the TALYS simulations.


\end{document}